\documentstyle[twocolumn,aps,prl,epsfig]{revtex}

\begin{document}

\title{\bf{Radiative Decay of the \boldmath $\psi(2S)$ into Two Pseudoscalar Mesons}}
\draft
\vspace{10mm}
% Following can be inserted into revtex document.

\author{
J.~Z.~Bai,$^1$   Y.~Ban,$^{11}$      J.~G.~Bian,$^1$
I.~Blum,$^{19}$  A.~D.~Chen,$^1$
G.~P.~Chen,$^1$  H.~F.~Chen,$^{18}$  
H.~S.~Chen,$^{1}$  J.~Chen,$^5$  
J.~C.~Chen,$^1$
% ...
\\ (BES Collaboration)}

\address{
$^1$Institute of High Energy Physics, Beijing 100039, People's Republic of
    China\\
% ...
$^5$Colorado State University, Fort Collins, Colorado 80523\\
% ...
$^{11}$Peking University, Beijing 100871, People's Republic of China\\
% ...
$^{18}$University of Science and Technology of China, Hefei 230026,
       People's Republic of China\\
$^{19}$University of Texas at Dallas, Richardson, Texas 75083-0688\\
% ...
}

%\begin{references}

%%\bibitem[\ddag]{atNU1} Present address:  Beijing University,
%% Beijing, China.

%\bibitem[\dagger]{support} Work supported in part by the National Natural
% Science Foundation of China under Contract No. 19991480, 19825116 and the
% Chinese Academy of Sciences under contract No. KJ95T-03,
% and by the Department of Energy under Contract Nos.
% DE-FG03-92ER40701 (Caltech), DE-FG03-93ER40788 (Colorado State University),
% DE-AC03-76SF00515 (SLAC), DE-FG03-91ER40679 (UC
% Irvine), DE-FG03-94ER40833 (U Hawaii), DE-FG03-95ER40925 (UT Dallas).

%\end{references}

\maketitle

\begin{abstract}
Radiative decays of the radially excited charmonium resonance, $\psi (2S),$
into $\pi \pi $, $K\overline{K}$ and $\eta \eta $ final states have been
measured in a sample of $3.96\times 10^{6}$ $\psi (2S)$ events
collected by the BES collaboration. The branching ratios
$B(\psi(2S)\rightarrow \gamma f_{2}(1270))=(2.27\pm 0.26\pm 0.39)\times
10^{-4}$ and
$B(\psi(2S)\rightarrow\gamma f_0(1710))\times B(f_0(1710)\rightarrow K^+K^-)
=(5.59\pm 1.12\pm 0.93)\times 10^{-5}$ are obtained.  When compared to 
the corresponding radiative $J/\psi$ decays, the observed $\psi(2S)$ 
radiative decay rates into $\gamma f_2(1270)$ and $\gamma f_0(1710)$ are 
consistent with the ``15\%'' rule.
\end{abstract}

% SECTION I

\section{Motivation}

In perturbative QCD, the dominant process of $J/\psi $ and $\psi (2S)$
hadronic decay is $c\overline{c}$ annihilation into three
gluons. Since the decay width is proportional to the amplitude of the $c%
\overline{c}$ wave function at the origin, $|\Psi (0)|^{2}$,
the branching fractions of $J/\psi $ and $\psi (2S)$ decays into light
quark states are related as \cite{15rule}:

\begin{eqnarray}
 & & \frac{B(\psi(2S)\rightarrow h)}{B(J/\psi\rightarrow h)} = \frac{%
B(\psi(2S)\rightarrow ggg)}{B(J/\psi\rightarrow ggg)}  \nonumber \\
&\simeq& \frac{B(\psi(2S)\rightarrow e^+ e^-)} {B(J/\psi\rightarrow e^+ e^-)}
= (14.6\pm2.2)\%  \nonumber
\end{eqnarray}

This relation is called \emph{the 15\% rule}. The
prediction was originally made for the total decay width into three gluons.
Since the partial widths of individual channels involving the initial
annihilation of $c\overline{c}$ quarks are also functions of the $|\Psi
(0)|^{2}$, we expect this rule to be generally valid.

Results from the Mark II experiment \cite{MarkIIresults} show that while
many of the $\psi(2S)$ hadronic decay channels obey this rule, it is
severely violated in vector plus pseudo-scalar (VP) final states such as $
\rho\pi$ and $K^*\overline{K}$ --- the so called $\rho\pi$ \textit{puzzle}.
The BES experiment also reported heavy suppression in the vector
plus tensor (VT) final states such as $K^{*}\overline{K^{*}_2}$, $\rho a_2$,
$\omega f_2$ and $\phi f_2^{^{\prime}}$ \cite{besVT}.
% as well as some of the
%axial-vector plus pseudo-scalar (AP) modes such as $K_1^{\pm}(1270)K^{\mp}$
%\cite{besAP}.

In perturbative QCD, the radiative $J/\psi $ and $\psi (2S)$ decays should
be similar to hadronic decays except instead of decaying into three gluons,
the radiative mode decays via two gluons and one photon.
Thus one power of the coefficient $\alpha _{S}$ is replaced by
$\alpha _{QED}$ in the cross section formula. It is expected that the
``15\%'' rule should also work for radiative decay modes \cite{Radiative15rule}.
Hence the ratio of $B(\psi(2S)\rightarrow \gamma X)$ to
$B(J/\psi \rightarrow \gamma X)$ for different final states $X$ should be
roughly 15\%. This paper explores the $\psi (2S)$ radiative decays into
pairs of pseudoscalars, $\pi\pi $, $K\overline{K}$ and $\eta \eta $, and
reports the first branching fraction measurements of $\psi(2S)\rightarrow
\gamma f_{2}(1270)$ and $\gamma f_0(1710)$. The branching fractions of
$\chi_{c0}$ and $\chi _{c2}$ decays into $\pi \pi $ and $\eta \eta $ are
also reported.

The $f_0(1710)$ has been observed with a large branching ratio in radiative
decays of $J/\psi$ into $K\overline{K}$, but not in the reaction $K^-p
\rightarrow KK\Lambda$ by the LASS experiment\cite{LASS}. The later
result excludes $f_0(1710)$ as a conventional $s\overline{s}$ state
and makes it a leading
glueball candidate. Thus whether $f_0(1710)$ can be seen in the radiative
$\psi(2S)$ decays is quite interesting.

% SECTION II

\section{BES Detector}

This study uses a subset of the 3.96 million $e^+e^-\rightarrow
\psi(2S)$ event\cite{BESpsipTotalNum} logged by the BES detector
operating at the BEPC storage ring. A detailed description of the BES detector
can be found elsewhere\cite{BESdetector}. It features a 40-layer main drift
chamber (MDC) in a 0.4 T solenoidal magnetic field providing a momentum
resolution of $\sigma _{p}/p=1.7\%\sqrt{1+p^{2}({\rm GeV/c)}}$ and a $dE/dx$
resolution of 9\% for hadron tracks. Outside of the MDC cylinder is an array
of 48 scintillation counters of the time-of-flight (TOF) system with a
resolution of 450 ps for hadrons. A 24-layer lead-gas barrel electromagnetic
shower calorimeter (BSC), outside of the TOF system, provides an energy
resolution of
$\sigma_{E}/E=0.22/\sqrt{E({\rm GeV})}$, and spatial resolutions of $\sigma _{\phi
}=4.5$ mrad and $\sigma _{\theta }=12$ mrad. The BSC is surrounded by
a magnetic coil and steel plates (for magnetic flux return). There is a
3-layer $ \mu $ counter interleaved with the steel plates to identify $\mu $
tracks.

The photons used in this analysis are detected as showers in
the BSC. Showers within ${\rm 8^{o}}$ of each other are regarded as split
showers of a single photon candidate and their energies are recombined.
For a $\pi^{0}$ or $\eta $ decaying
into two photons, the helicity angle of the decay should be flat and hence
$|\cos\theta _{helicity}|<0.99$ is required to remove some asymmetric
background decays. For charged tracks, only those that fall
in the fiducial region $|\cos\theta |<0.8$
are used. Charged tracks identified as electrons or positrons by the BSC
or identified as $ \mu ^{+}$ or $\mu ^{-}$ by the $\mu $ counter are
rejected. TOF information, $ dE/dx$ information and kinematic fitting are
used to identify charged particles. Kinematic fits are applied to
improve momentum measurements and mass resolutions, as well as to resolve
combinatorial ambiguities if more than one combination is possible
in the same event.

% SECTION III

\section{\boldmath $\psi(2s)\rightarrow\gamma\pi\pi$ Analysis}

\begin{figure}
 \centerline{\epsfig{file=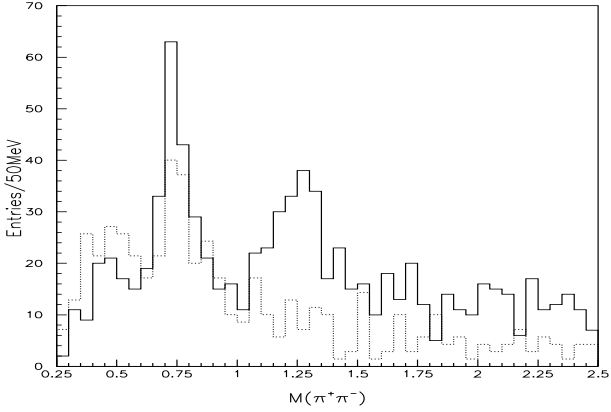,width=9cm,height=6cm}}
 \caption{\label{fig:MpipiForPsipTau}
          $M_{\pi^+\pi^-}$ from $\psi(2S)$ data in the solid line histogram
          and luminosity normalized $\tau$ data in the dashed line histogram.}
\end{figure}

\begin{figure}
 \centerline{\epsfig{file=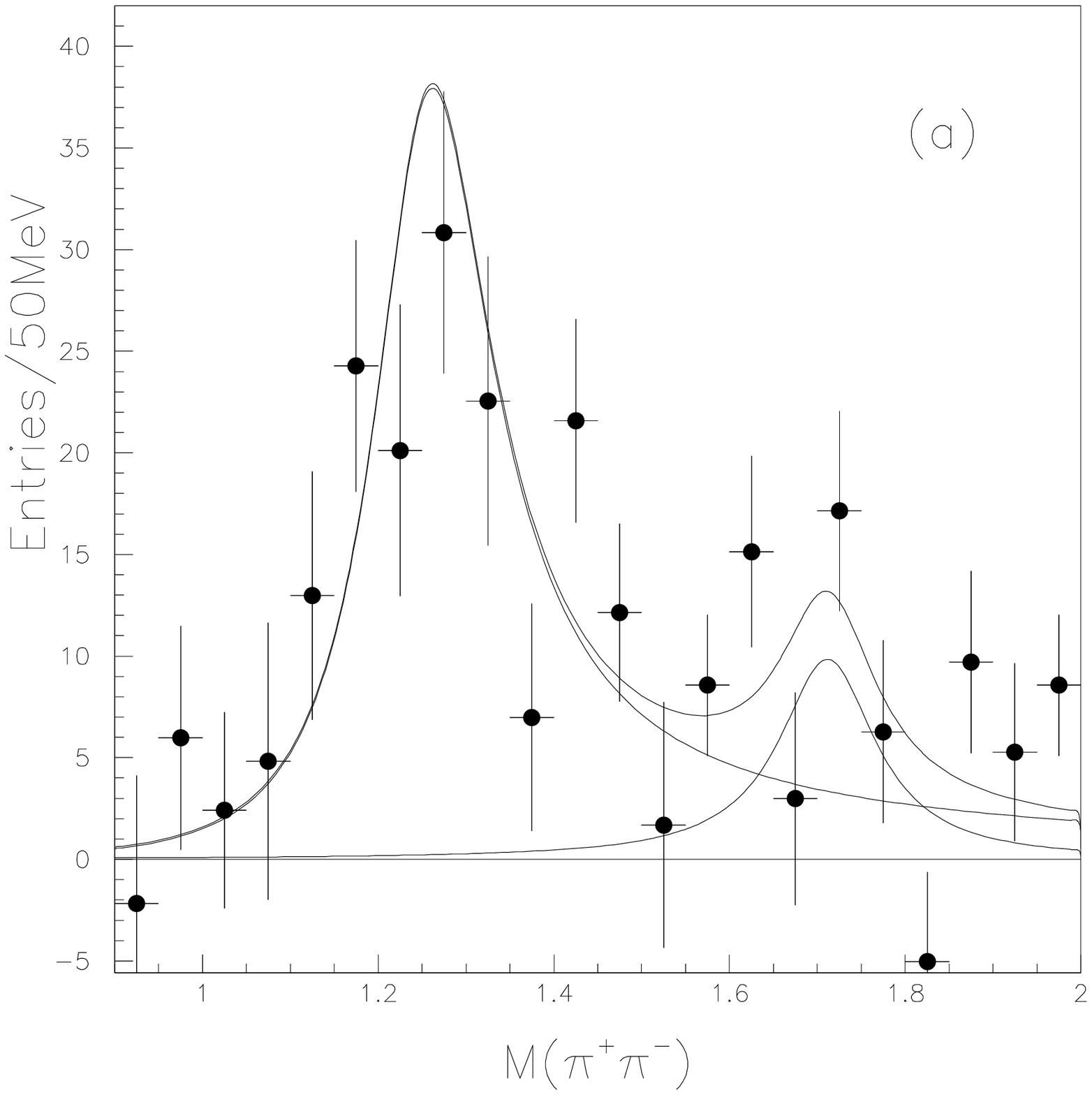,width=4cm,height=6cm}
             \epsfig{file=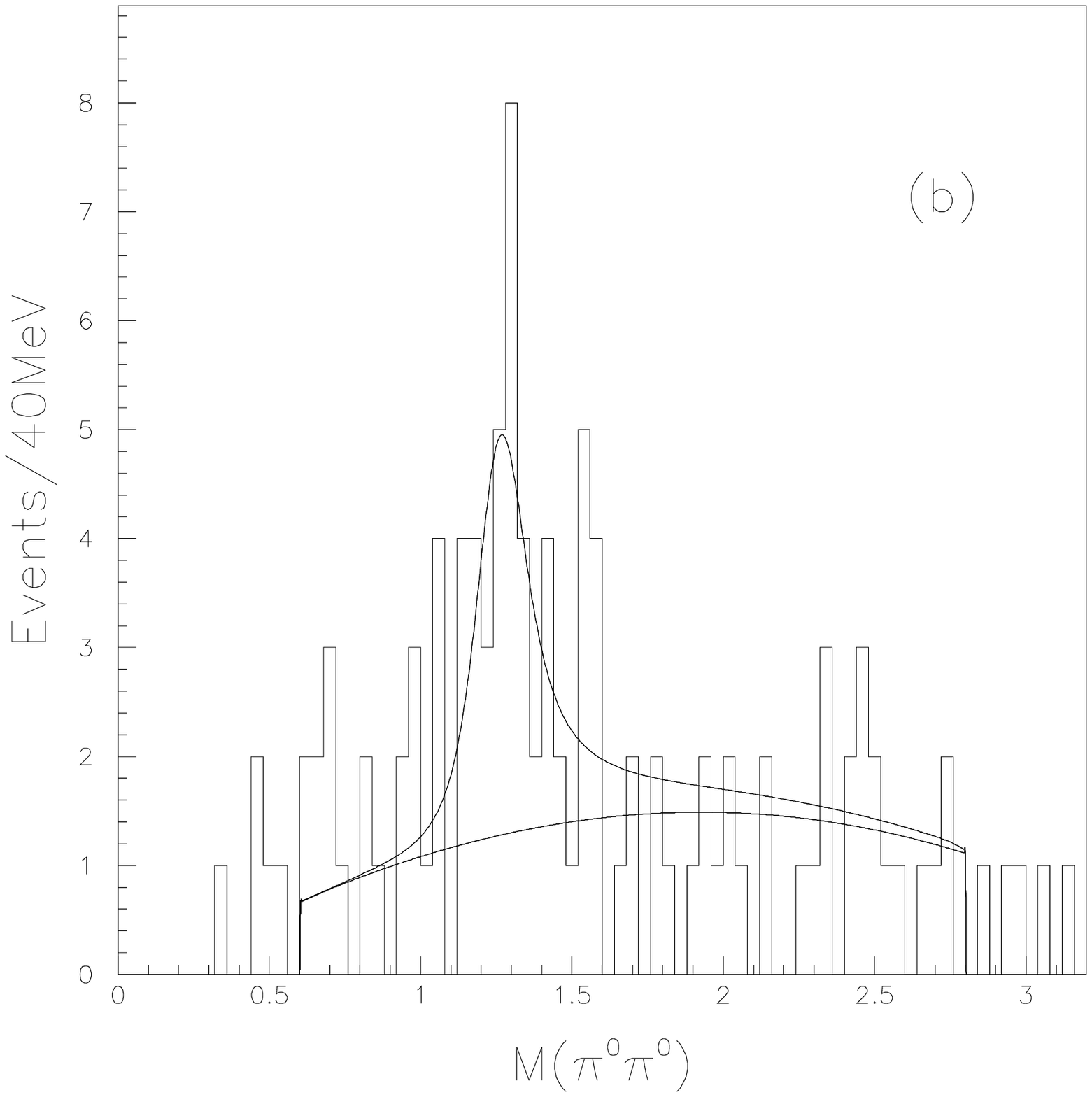,width=4cm,height=6cm}}
 \caption{\label{fig:Mpipi}
          (a): $M_{\pi^+\pi^-}$ fit result. The data points are obtained from 
               the difference of the two histograms in Fig.
               \ref{fig:MpipiForPsipTau}             
          (b): $M_{\pi^0\pi^0}$.}
\end{figure}

Here the analysis of decays into both charged and neutral pion pairs,
$\psi (2S)\rightarrow \gamma \pi ^{+} \pi ^{-}$ and
$\psi (2S)\rightarrow \gamma \pi ^{0}\pi ^{0}$, is described.
In the charged channel, events
are selected with two oppositely charged tracks and at least one photon.
Kinematic fit, TOF, and $dE/dx$ information are combined and the probability
for the two pions hypothesis should be greater than 1\% and should also be
greater than the probability for the two kaons hypothesis. The solid-line
histogram of Fig. \ref{fig:MpipiForPsipTau} is the $\pi^{+}\pi ^{-}$
invariant mass distribution below 2.5 GeV. A clear $\rho $ signal is
observed over a continuous background. These are due to the initial state
radiation processes $e^{+}e^{-}\rightarrow \gamma \rho $ and
$e^{+}e^{-}\rightarrow \gamma \mu ^{+}\mu ^{-}$. These processes
are also presented in the $e^{+}e^{-}\rightarrow \tau \tau$ scan data taken
by BES at $\sqrt{s}=3.55 - 3.6$ GeV (below $\psi(2S)$ threshold)
\cite{tautau} and the mass distribution from the $\tau \tau$ scan can be
used to represent the background in the $\psi(2S)$ sample.
The $\pi^{+}\pi ^{-}$ mass distribution taken from the $\tau$ scan data is shown
in Fig. \ref{fig:MpipiForPsipTau} (with the dashed line).
The backgrounds are removed by assigning
each $\psi (2S)$ event a weight of 1 and each
$\tau $ event a weight of $-w$ in the likelihood function for the mass fit.
Here $w$ is the ratio of the integrated luminosities of these two data sets.
Fig. \ref {fig:Mpipi}a shows the result of subtracting the $\tau \tau$ scan 
histogram normalized to the $\psi(2S)$ from the $\psi(2S)$ histogram.

%Figure \ref{fig:Mpipi}b is the result of the second channel, $\psi (2S)
%\rightarrow \gamma \pi ^{0}\pi ^{0}\rightarrow \gamma \gamma \gamma \gamma
%\gamma $.

In the neutral mode, events are required to have at least 5 neutral
tracks and no
charged tracks. A six-constraint fit is made to all possible $\gamma \pi^0
\pi^0$ combinations with two $\pi^0$ resonances. The combination
with the smallest fit $\chi ^{2}$ is selected. A four-constraint fit
is also applied on that combination and $|M_{\gamma \gamma }-M_{\pi^0}|
< 70MeV$ is required. The resulting mass distribution is shown in
Figure \ref{fig:Mpipi}b.

In both the $\pi^+ \pi^-$ and $\pi^0 \pi^0$ invariant mass distributions,
clear $f_{2}(1270)$ signals are observed. Both distributions are fitted with
a D-wave Breit-Wigner function with the resonant parameters fixed at the
PDG \cite{PDG2000} values for the $f_2(1270)$. An S-wave Breit-Wigner with mass and width fixed
at the PDG values for the $f_0(1710)$ is included in the mass fitting
in the $\gamma \pi ^{+}\pi ^{-}$ channel in order to describe the line shape
in that region. A phase space background is included in the $\gamma\pi^0\pi^0$
channel. The fit yields $209.8\pm 24.9$ events and $29.9\pm 11.1$ events above
background as shown in Fig. \ref{fig:Mpipi}a and Fig. \ref{fig:Mpipi}b,
respectively.

Branching fractions of $B(\psi(2S)\rightarrow\gamma f_{2}(1270))=(2.21\pm
0.26\pm 0.39) \times 10^{-4}$ from the $\psi(2S)\rightarrow \gamma
\pi^+\pi^-$ channel and $B(\psi(2S)\rightarrow \gamma f_{2}(1270))=(2.95\pm
1.10\pm 1.12)\times 10^{-4}$ from the $\psi(2S)\rightarrow \gamma\pi^0\pi^0$
channel are obtained using the total number of $\psi (2S)$ events and
detection efficiencies, which are determined from Monte Carlo simulations, of
43.8\% and 9.6\%, respectively. The combined result from these two channels is
\[
B(\psi(2S)\rightarrow\gamma f_{2}(1270))=(2.27\pm 0.26\pm 0.39)\times 10^{-4}.
\]
Here and below, the first error is statistical and the second is systematic. The latter
includes uncertainties from varying the cuts, the shape of the background,
and the uncertainty from the total number of $\psi(2S)$'s. This result is
consistent with the ``15\%'' rule when compared with the corresponding
branching fraction from $J/\psi $ decay on PDG. See Table 
\ref{tab:15rule}.

\begin{figure}
 \centerline{\epsfig{file=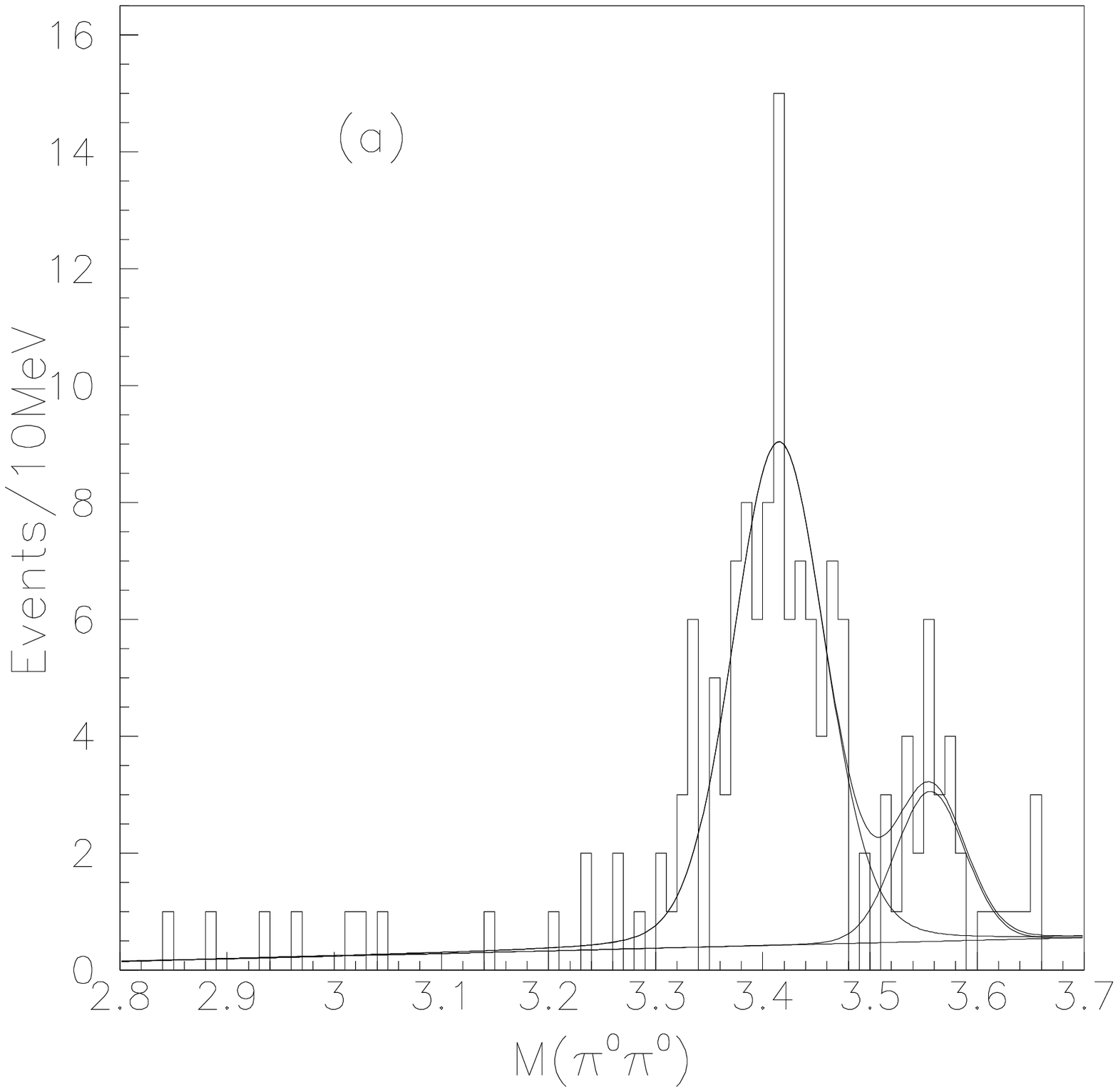,width=4cm,height=6cm}
             \epsfig{file=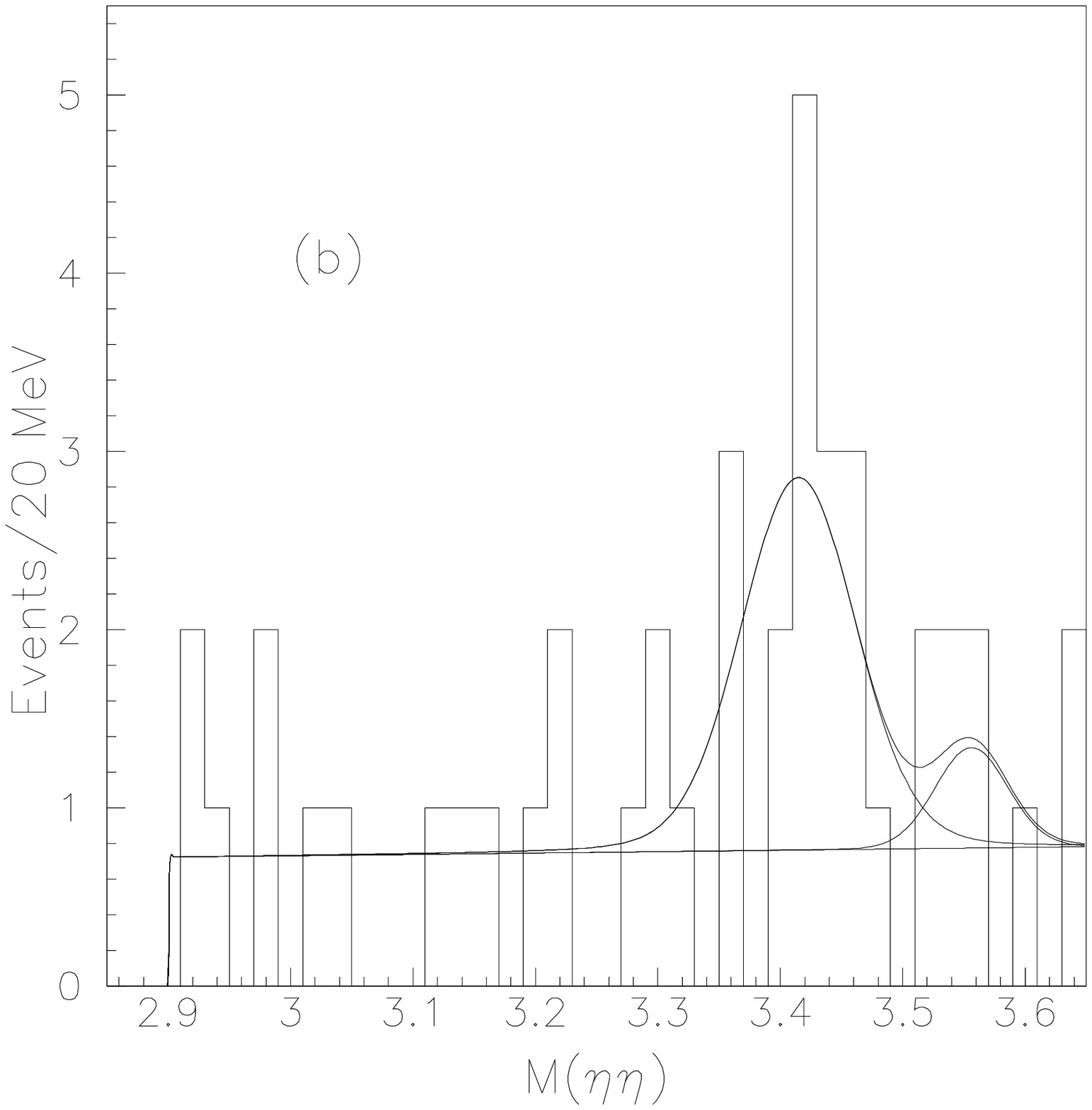,width=4cm,height=6cm}}
 \caption{\label{fig:Mgammas}
          Invariant mass of (a) $\pi^0\pi^0$ and (b) $\eta\eta$.}
\end{figure}

A $f_0(1710)$ signal is observed in the $\pi^+\pi^-$ invariant mass distribution, as
shown in Fig. \ref{fig:Mpipi}a. The number of $f_0(1710)$ events above
background is $39.2\pm 10.1$. The Monte Carlo determined efficiency for this
channel is 45.4\%, and the resulting branching fractions are
\begin{eqnarray*}
B(\psi(2S)\rightarrow\gamma f_0(1710)) & \times &
B(f_0(1710)\rightarrow \pi\pi)  \\
           & = & (3.38\pm 0.87\pm1.41)\times 10^{-5} 
\end{eqnarray*}
or
\[
           < 5.50\times 10^{-5}\ \ (90\%\ C.L.\cite{UpperLimit})
\]
The region with a $\pi^+\pi^-$ invariant mass greater than 3 GeV has been
presented elsewhere \cite{BESchi_c}. The region with a $\pi^0\pi^0$ invariant
mass greater than 3 GeV (see Fig. \ref{fig:Mgammas}a) has signal peaks
due to the $\chi _{c0}$ and $\chi _{c2}$ charmonium states. This mass
distribution is fitted with two Breit-Wigner resonancess plus a polynomial 
background
function, and $96.9\pm 11.1$ and $20.8\pm 5.8$ events are obtained for
$\chi _{c0}$ and $\chi _{c2}$, respectively. The detection efficiencies are
$10.5\%$ and $8.2\%$, and the resulting branching fractions are

\[
B(\chi _{c0}\rightarrow \pi ^{0}\pi ^{0})=(2.65\pm 0.30\pm 0.58)\times 10^{-3}
\]
and
\[
B(\chi _{c2}\rightarrow \pi ^{0}\pi ^{0})=(8.7\pm 2.4\pm 5.0)\times 10^{-4}
\]
The detection efficiencies, branching fraction acceptances, and final
results for the decays described in this section are summarized in Tables
\ref{tab:bf_ratios_11} to \ref{tab:bf_ratios_22}.

% SECTION IV

\section{\boldmath $\psi(2s)\rightarrow\gamma K\overline{K}$ Analysis}

\begin{figure}
 \centerline{\epsfig{file=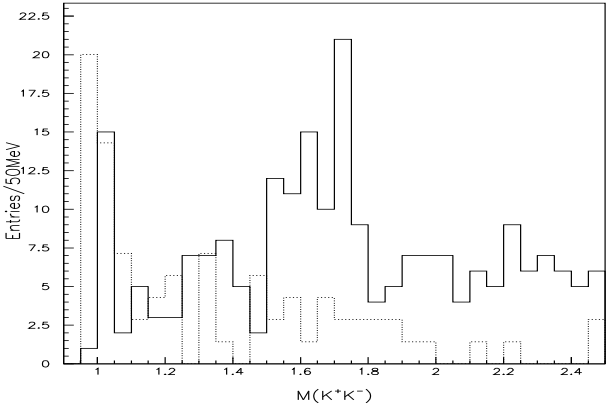,width=9cm,height=6cm}}
 \caption{\label{fig:MKKForPsipTau}
          $M_{K^+K^-}$ from $\psi(2S)$ data in the solid line histogram
          and luminosity normalized $\tau$ data in the dashed line histogram.}
\end{figure}

\begin{figure}
  \centerline{\epsfig{file=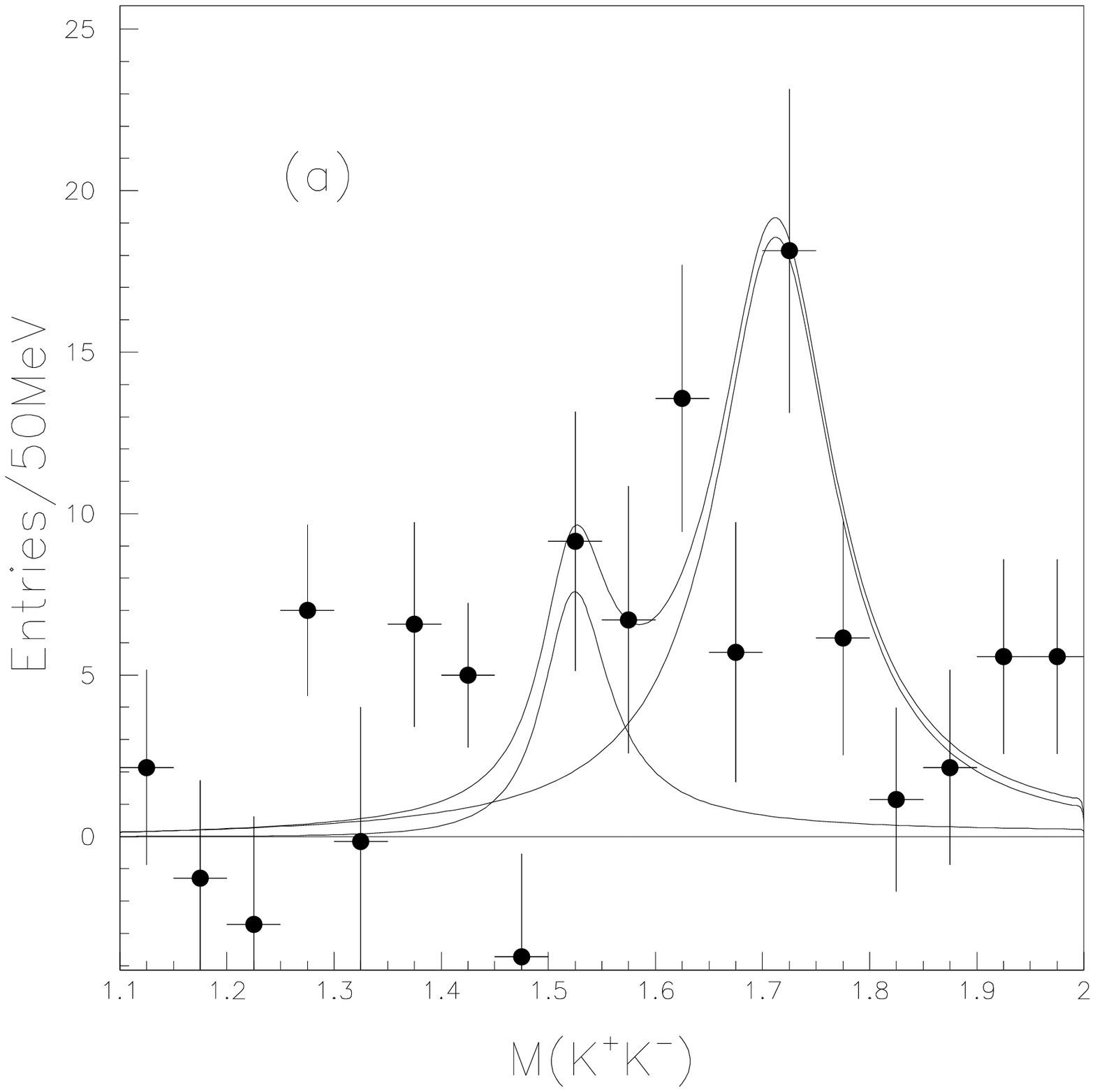,width=4cm,height=6cm}
              \epsfig{file=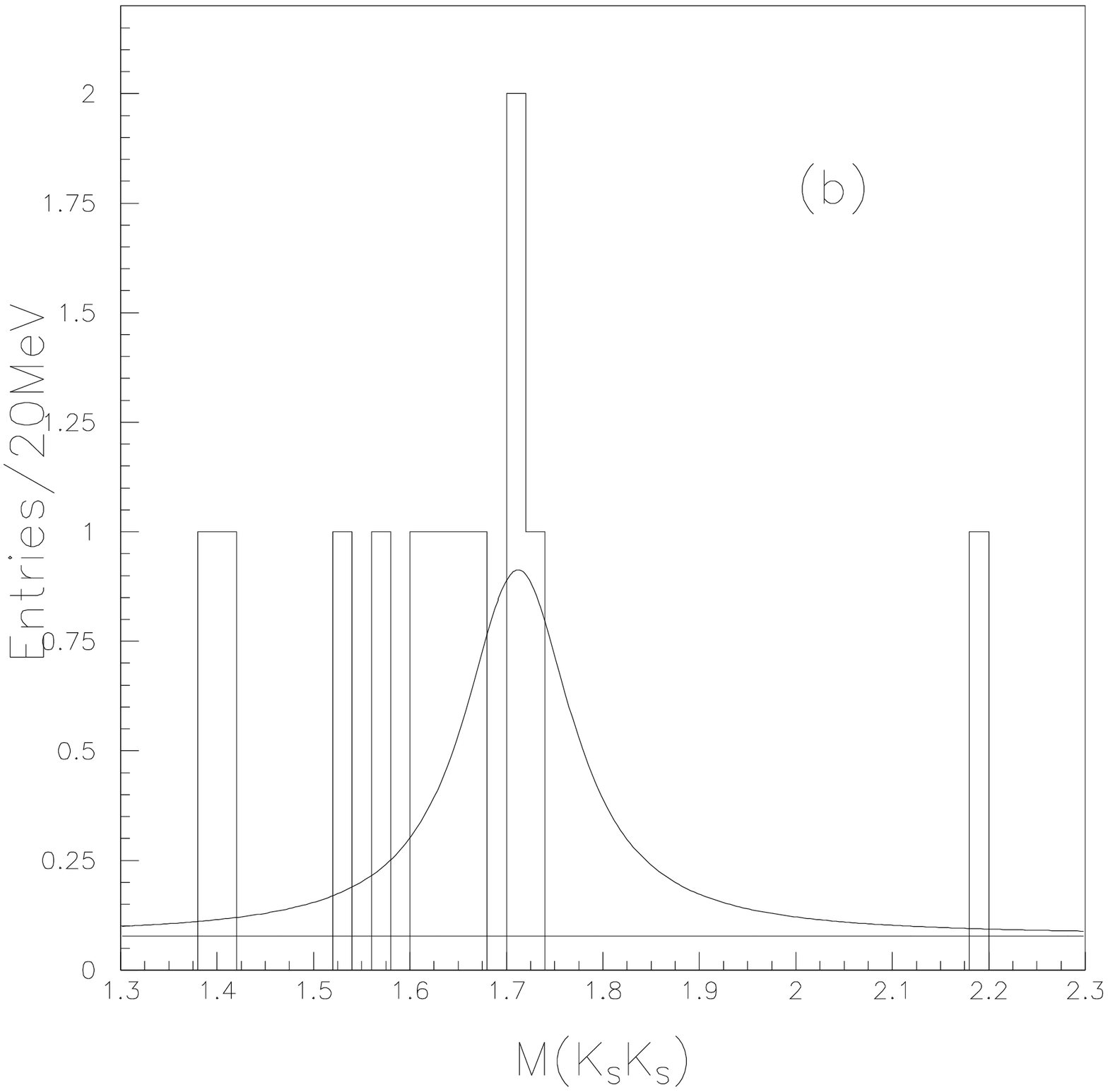,width=4cm,height=6cm}}
  \caption{\label{fig:Mkk}
           Invariant mass of (a) $K^+K^-$ and (b) $K_S^0K_S^0$. The data
           points in (a) are obtained from the difference of the two 
           histograms in Fig. \ref{fig:MKKForPsipTau}.}
\end{figure}

Here the analysis of decays into charged and neutral kaon pairs,
$\psi(2S)\rightarrow\gamma K^+K^-$ and $\psi (2S)\rightarrow \gamma
K_{S}^{0}K_{S}^{0}\rightarrow \gamma 4\pi^{\pm}$, is presented. For the
$\psi(2S)\rightarrow\gamma K^+K^-$ channel, cuts similar to those for
the $\gamma\pi^+ \pi^-$ analysis are used, but with the requirement
that the probability from particle identification for each kaon
hypothesis should be greater than 0.01 and should be greater than the
probability for the pion hypothesis. QED backgrounds such as
$e^+e^-\rightarrow\gamma\phi\rightarrow \gamma K^+K^-$ and $e^+e^-
\rightarrow\gamma\mu^+\mu^-$ are determined using the $\tau$ scan data (See
Fig.  \ref{fig:MKKForPsipTau}).  A $f_0(1710)$ signal and a hint of a
possible $f_2^{^{\prime }}(1525)$ signal (Fig. \ref{fig:Mkk}) are
observed. The mass distribution is fitted using S-wave
and D-wave Breit-Wigner functions with masses and widths fixed at the PDG values
for $f_0(1710)$ and $f_{2}^{^{\prime }}(1525)$, respectively. The fit
yields $71.9\pm 14.4$ $f_0(1710)$ events above background. The
detection efficiency is $33.6\%$, giving a branching fraction

\begin{eqnarray*}
  B(\psi (2S) \rightarrow \gamma f_0(1710)) & \times &
  B(f_0(1710) \rightarrow K^+K^-)  \\
              & = & (5.59\pm 1.12\pm 0.93)\times 10^{-5} 
\end{eqnarray*}
The systematic error
includes the uncertainties from varying the cuts and the total number
of $\psi(2S)$'s. This result is again consistent with the ``15\%'' rule.
See Table \ref{tab:15rule}.

For the $\psi (2S)\rightarrow \gamma K_{S}^{0}K_{S}^{0}\rightarrow \gamma
\pi ^{+}\pi ^{-}\pi ^{+}\pi ^{-}$ channel, events with two positive charged
tracks, two negative charged tracks and at least one photon are selected.
Both $K_{S}^{0}$ vertices are reconstructed and $|M_{\pi^+\pi^-}-M_{K_S^0}|$
must be smaller than 20 $MeV$. At least one of the combinations must
yield a four-constraint fit with $\chi^2$ probability greater than 0.01. If
more than one combination survives, the combination with the smallest
value of
\[
\sqrt{(m_{\pi _{1}^{+}\pi _{2}^{-}}-m_{K_{S}^{0}})^{2}+(m_{\pi _{3}^{+}\pi
_{4}^{-}}-m_{K_{S}^{0}})^{2}}
\]
is chosen. An S-wave Breit Wigner plus a polynomial background are used to fit
the invariant mass distribution shown in Fig. \ref{fig:Mkk}b. The fit
finds $6.8\pm 3.1$ events, or an upper
limit of 10.8 events at 90\% confidence level. The detection
efficiency for this channel is $18.0\%$. The branching fraction is
\begin{eqnarray*}
  B(\psi (2S)\rightarrow \gamma f_0(1710)) & \times &
  B(f_0(1710)\rightarrow K^0_SK^0_S) \\
  & = & (2.10\pm 0.96 \pm 1.11)\times 10^{-5}
\end{eqnarray*}
or
\[
  <3.98\times 10^{-5}\ \ (90\%\ C.L.)
\]
The detection efficiencies, branching fraction acceptances, and final
results for the decays described in this section are summarized in Tables
\ref{tab:bf_ratios_11} to \ref{tab:bf_ratios_22}.

% SECTION V

\section{\boldmath $\psi(2s)\rightarrow\gamma\eta\eta$ Analysis}
Here
the analysis of decays into $\eta$ pairs, $\psi(2S)\rightarrow\gamma
\eta\eta\rightarrow 5\gamma$, is presented. The selection criteria for this
channel are similar to those used in the $\psi(2S)\rightarrow\gamma
\pi^0\pi^0$ channel except that the requirement $|M_{\gamma \gamma}-M_{\eta }|
< 70$ $MeV$ is imposed. In the region where $M_{\eta\eta}
<3$ GeV, no $\eta \eta $ resonant state is observed. In the mass region with
$M_{\eta\eta }>3 $ GeV, shown in Fig. \ref{fig:Mgammas}b, a $\chi _{c0}$ signal
and a possible $ \chi _{c2}$ signal are observed. Two Breit-Wigner functions
with a polynomial background are used to fit the mass distribution, and
$12.7\pm 5.3$ events for $\chi _{c0}$ and an upper limit of 5.9 $\chi _{c2}$
events are found, giving the following branching fractions
\[
B(\chi _{c0}\rightarrow \eta \eta )=(1.94\pm 0.81\pm 0.59)\times 10^{-3}
\]
\[
B(\chi _{c2}\rightarrow \eta \eta )<1.22\times 10^{-3}\ \ (90\%\ C.L.)
\]
The detection efficiencies for these two channels are $11.9\%$ and $10.5\%$,
respectively.

Flavor SU(3) symmetry predicts that the branching fractions of $\chi _{c0}$
decay into $\pi ^{0}\pi ^{0}$ and $\eta \eta $ should be the same except
for a phase space factor and a barrier factor of $p^{(2s+1)}$, where
$p$ is the momentum of the $\pi ^{0}$ or $\eta $ in $\chi _{c}$'s rest frame
and $s$ is the spin of the $\chi _{c}$. Based on the PDG values for the
$\chi _{c0}$, this predicts $B(\chi_{c0}\rightarrow\eta\eta )/B(\chi_{c0}
\rightarrow \pi^0\pi^0)=0.95$ which is consistent with our measurement of
\[
\frac{B(\chi _{c0}\rightarrow \eta \eta )}{B(\chi _{c0}\rightarrow \pi
^{0}\pi ^{0})}=0.73\pm 0.32\pm 0.27 .
\]
The detection efficiencies, branching fraction acceptances, and final
results for the decays described in this section are summarized in Tables
\ref{tab:bf_ratios_11} to \ref{tab:bf_ratios_22}.

% SECTION VI

\section{\boldmath $\psi(2S)$ Normalization}

The total number of $\psi(2S)$ events in the BES data sample is determined from
the observed number of cascade decays of $\psi(2S)\rightarrow\pi^+\pi^-
J/\psi$, $J/\psi\rightarrow X$.
The total number of $\psi(2S)\rightarrow \pi^+\pi^-J/\psi$ events
observed in the full BES data sample is
$(1.227\pm 0.003\pm 0.017)
\times 10^6$ \cite{BESpsipTotalNum} \cite{BESpsipTotalNumPapar}.
The decay branching fraction of
$B(\psi(2S)\rightarrow\pi^+\pi^- J/\psi)=(31.0\pm 2.8)\%$
taken from PDG 2000\cite{PDG2000}
is used to calculate the total number of produced
$\psi(2S)$ events in the data sample.
This paper provides the ratios of
the branching fractions and $B(\psi(2S)\rightarrow\pi^+
\pi^-J/\psi)$ in order to isolate the systematic error caused by the
$\psi(2S)\rightarrow\pi^+\pi^-J/\psi$ branching fraction. See Tables
\ref{tab:bf_ratios_11} to \ref{tab:bf_ratios_22} for the results.

% SECTION VII

\section{Summary}

This paper studies $\psi(2S) \rightarrow \gamma \pi^+\pi^-$, $\gamma \pi^0
\pi^0$, $\gamma K^+K^-$, $\gamma K^0_S K^0_S$, $\gamma \eta\eta$ final states
and reports the first 
measurement of the $\psi(2S)\rightarrow \gamma f_{2}(1270)$ and $\psi 
(2S)\rightarrow \gamma f_0(1710)\rightarrow \gamma K^+K^-$ and $\gamma 
K_S^0 K_S^0$ branching fractions. 
A clear $f_0(1710)$ signal in $\psi(2S)$ radiative decay into $K^+K^-$
final states is observed. 
The results are consistent with the ``15\%'' rule.

In addition, this paper reports the first measurement of the
branching fractions of $\chi_{c0}$ and $\chi_{c2}$ decay into $\pi^0\pi^0$,
$\chi_{c0}$ decay into $\eta\eta$, and an upper limit of the branching
fraction of $\chi_{c2}$ decay into $\eta\eta $. The results from $\chi_{c0}
\rightarrow \pi^0\pi^0$ and $\eta\eta$ are consistent with the prediction
by SU(3) flavor symmetry.

%\SECTION VIII

\section{Acknowledgements}

We acknowledge the strong support from the BEPC accelerator staff and the
IHEP computer center. The work is supported in part by the National Natural
Science Foundation of China under Contract No. 19290400 and the Chinese
Academy of Sciences under contract No. H-10 and E-01 (IHEP), and by the
Department of Energy under Contract No. DE-FG03-92ER40701 (Caltech),
DE-FG03-93ER40788 (Colorado State University), DE-AC03-76SF00515 (SLAC),
DE-FG03-91ER40679 (UC Irvine), DE-FG03-94ER40833 (U Hawaii),
DE-FG03-95ER40925 (UT Dallas).

\onecolumn

\begin{table}[tb] \centering
 \begin{tabular}{|l|l|l|l|}   \hline
 Final state & 
               $B(\psi(2S)\rightarrow)(\times 10^{-4})$ &
               $B(J/\psi\rightarrow)(\times 10^{-4})$ & 
               $B(\psi(2S))/B(J/\psi)$ \\ \hline\hline
 $\gamma f_2(1270)$ &
               $2.27\pm 0.26\pm 0.39$ &
               $13.8\pm 1.4$ & 
               $(16.4\pm 3.1)$\% \\ \hline
 $\gamma f_0(1710)\rightarrow\gamma K^+K^-$ &
               $0.559\pm 0.112\pm 0.093$ &
               $4.25^{+0.60}_{-0.45}$ \cite{PDG2000} &
               $(13.2^{+3.8}_{-4.0})$\% \\ \hline
 \end{tabular}
 \caption{Verification of the 15\% rule. The value of $B(J/\psi
          \rightarrow \gamma f_0(1710)\rightarrow\gamma K^+K^-)$ is 
          obtained from $B(J/\psi\rightarrow\gamma f_0(1710)\rightarrow
          \gamma K \overline{K})=8.6^{+1.2}_{-0.9}$ 
          divided by a factor of 2 to account for isospin.}
 \label{tab:15rule}
\end{table}

\begin{table}[tb] \centering
 \begin{tabular}{|l|l|l|l|l|}   \hline
  Mode & Number of events & Detection & Branching Fractions & Number of events\\
       & from mass fitting & efficiency & correction factor & after correction
                                                               \\ \hline\hline
  $\psi(2S)\rightarrow\gamma f_2(1270)$ from $\gamma\pi^+\pi^-$ & 
    $209.8\pm 24.9$ & $43.8\%$ & $0.847\pm 0.024$ & $565.5\pm 67.1\pm 85.7$
                                                                      \\ \hline
  $\psi(2S)\rightarrow\gamma f_2(1270)$ from $\gamma\pi^0\pi^0$ & 
    $29.9\pm 11.1$ & $9.6\%$ & $0.827\pm 0.028$ & $376.7\pm 139.9\pm 138.1$
                                                                      \\ \hline
  $\psi(2S)\rightarrow\gamma f_0(1710)\rightarrow\gamma\pi\pi$ 
                                    from $\gamma\pi^+\pi^-$ &
    $39.2\pm 10.1$ & $45.4\%$ & $1.0\pm 0.0$ & $86.3\pm 22.2\pm 35.2$ \\ \hline 
  $\psi(2S)\rightarrow\gamma f_0(1710)\rightarrow\gamma K^+K^-$ &
    $71.9\pm 14.4$ & $33.6\%$ & $1.0\pm 0.0$ & $214.0\pm 42.9\pm 30.6$
                                                                      \\ \hline
  $\psi(2S)\rightarrow\gamma f_0(1710)\rightarrow\gamma K^0_SK^0_S$ & 
    $6.8\pm 3.1$ & $18.0\%$ & $0.471\pm 0.004$ & $80.3\pm 36.6\pm 41.9$ 
                                                                      \\ \hline
 \end{tabular}
 \caption{Numbers of events before and after correction for efficiency and 
          branching fraction acceptance. Branching fraction acceptance factors
          include branching fractions from intermediate decays processes such 
          as $\pi^0\rightarrow\gamma\gamma$, $\eta\rightarrow\gamma\gamma$, 
          $K^0_S\rightarrow\pi^+\pi^-$, $f_2(1270)\rightarrow\pi\pi$, etc.
          They do not include iso-spin factors.} 
 \label{tab:bf_ratios_11}
\end{table}

\begin{table}[tb] \centering
 \begin{tabular}{|l|l|l|} \hline
  Mode & $B(\times 10^{-4})$ & $B/B(\psi(2S)\rightarrow
                      \pi^+\pi^-J/\psi) (\times 10^{-4})$ \\ \hline\hline
  $\psi(2S)\rightarrow\gamma f_2(1270)$ from $\gamma\pi^+\pi^-$ &
            $2.21\pm 0.26\pm 0.39$ & $7.14\pm 0.85\pm 1.09$ \\ \hline
  $\psi(2S)\rightarrow\gamma f_2(1270)$ from $\gamma\pi^0\pi^0$ &
            $2.95\pm 1.10\pm 1.12$ & $9.52\pm 3.53\pm 3.49$ \\ \hline
  $\psi(2S)\rightarrow\gamma f_2(1270)$ from $\gamma\pi\pi$ &
            $2.27\pm 0.26\pm 0.39$ & $7.31\pm 0.83\pm 1.04$ \\ \hline
  $\psi(2S)\rightarrow\gamma f_0(1710)\rightarrow\gamma\pi\pi$ from 
                                              $\gamma\pi^+\pi^-$ &
            $0.338\pm 0.087\pm 0.141$ & $1.09\pm 0.28\pm 0.44$ \\ \hline
  $\psi(2S)\rightarrow\gamma f_0(1710)\rightarrow\gamma K^+K^-$ &
            $0.559\pm 0.112\pm 0.093$ & $1.80\pm 0.36\pm 0.26$ \\ \hline
  $\psi(2S)\rightarrow\gamma f_0(1710)\rightarrow\gamma K^0_SK^0_S$ &
            $0.210\pm 0.096\pm 0.111$ & $0.68\pm 0.31\pm 0.35$ \\ \hline
 \end{tabular}
 \caption{Branching fractions and ratios of branching
          fractions ($B/B(\psi(2S)\rightarrow\pi^+\pi^-J/\psi)$) for
          $\psi(2S)\rightarrow\gamma X\rightarrow\gamma P\overline{P}$
          modes ($P$ stands for pseudo-scalar). }
 \label{tab:bf_ratios_12}
\end{table}

\begin{table}[tb] \centering
 \begin{tabular}{|l|l|l|l|l|}   \hline
  Mode & Number of event & Detection & Branching Fractions & Number of events\\
       & from mass fitting & efficiency & correction factor & after correction
                                                               \\ \hline\hline
  $\chi_{c0}\rightarrow\pi^0\pi^0$ & 
    $96.9\pm 11.1$ & $10.5\%$ & $0.9761\pm 0.0005$ & $945.4\pm 108.3\pm 163.8$ 
                                                               \\ \hline
  $\chi_{c2}\rightarrow\pi^0\pi^0$ & 
    $20.8\pm 5.8$ & $8.2\%$ & $0.9761\pm 0.0005$ & $259.9\pm 72.5\pm 145.8$ 
                                                               \\ \hline
  $\chi_{c0}\rightarrow\eta\eta$ & 
    $12.7\pm 5.3$ & $11.9\%$ & $0.1547\pm 0.0035$ & $689.9\pm 287.9\pm 188.9$
                                                               \\ \hline
  $\chi_{c2}\rightarrow\eta\eta$ &
    $<5.9$ & $10.5\%$ & $0.1547$ & $<363.3$                    \\ \hline
 \end{tabular}
 \caption{Numbers of events corrected for efficiency and branching fraction
          acceptance for $\chi_c$ decay.}
 \label{tab:bf_ratios_21}
\end{table}

\begin{table}[tb] \centering
 \begin{tabular}{|l|l|l|l|}   \hline
  Mode & $B(\times 10^{-3})$ & 
  $B\times B(\psi(2S)\rightarrow\gamma\chi_{c0,2}) (\times 10^{-4})$ & 
  $B\times B(\psi(2S)\rightarrow\gamma\chi_{c0,2})/
   B(\psi(2S)\rightarrow\pi^+\pi^-J/\psi) (\times 10^{-4})$ \\ \hline\hline
  $\chi_{c0}\rightarrow\pi^0\pi^0$ & 
    $2.65\pm 0.30\pm 0.58$ & $2.47\pm 0.28\pm 0.49$ & $7.96\pm 0.91\pm 1.38$
                                                                   \\ \hline
  $\chi_{c2}\rightarrow\pi^0\pi^0$ &
    $0.87\pm 0.24\pm 0.50$ & $0.68\pm 0.19\pm 0.39$ & $2.19\pm 0.61\pm 1.23$
                                                                   \\ \hline
  $\chi_{c0}\rightarrow\eta\eta$ & 
    $1.94\pm 0.81\pm 0.59$ & $1.80\pm 0.75\pm 0.52$ & $5.81\pm 2.42\pm 1.59$
                                                                   \\ \hline
  $\chi_{c2}\rightarrow\eta\eta$ & 
    $<1.22$                & $<0.95$                & $<3.06$      \\ \hline
 \end{tabular}
 \caption{The $\chi_c$ decay branching fractions, and ratios of branching 
          fractions for $\chi_{c0,2}\rightarrow\pi^0\pi^0$ or $\eta\eta$
          ($B\times B(\psi(2S)\rightarrow\gamma \chi_{c0,2})/
                      B(\psi(2S)\rightarrow\pi^+\pi^-J/\psi)$). }
 \label{tab:bf_ratios_22}
\end{table}

\end{document}